\documentclass{emulateapj}
\newcommand{\spit}{\textit{Spitzer Space Telescope}}

\newcommand{\rs}{0.928}
\newcommand{\ersp}{+0.018}
\newcommand{\ersm}{-0.013}

\newcommand{\rp}{1.184}
\newcommand{\erpp}{+0.028}
\newcommand{\erpm}{-0.018}

\newcommand{\semi}{0.0488}
\newcommand{\esemi}{0.0005}

\newcommand{\flone}{0.00086}
\newcommand{\eflone}{0.00007}
\newcommand{\hjdone}{2454323.28342}
\newcommand{\ehjdone}{0.00242}
\newcommand{\dtone}{ -5.42}
\newcommand{\edtone}{3.49}
\newcommand{\chione}{ 0.76}
       
\newcommand{\fltwo}{0.00122}
\newcommand{\efltwo}{0.00009}
\newcommand{\hjdtwo}{2454193.21175}
\newcommand{\ehjdtwo}{0.00259}
\newcommand{\dttwo}{ -6.94}
\newcommand{\edttwo}{3.72}
\newcommand{\chitwo}{ 0.84}
       
\newcommand{\flthr}{0.00261}
\newcommand{\eflthr}{0.00031}
\newcommand{\hjdthr}{2454323.29066}
\newcommand{\ehjdthr}{0.00488}
\newcommand{\dtthr}{  5.00}
\newcommand{\edtthr}{7.03}
\newcommand{\chithr}{ 0.89}
       
\newcommand{\flfour}{0.00210}
\newcommand{\eflfour}{0.00029}
\newcommand{\hjdfour}{2454193.22043}
\newcommand{\ehjdfour}{0.00486}
\newcommand{\dtfour}{  5.56}
\newcommand{\edtfour}{7.00}
\newcommand{\chifour}{ 0.99}

\newcommand{\hjdaug}{2454323.28485}
\newcommand{\ehjdaug}{0.00217}
\newcommand{\hjdaugm}{-3.4}
\newcommand{\ehjdaugm}{ 3.1}

\newcommand{\hjdapr}{2454193.21366}
\newcommand{\ehjdapr}{0.00228}
\newcommand{\hjdaprm}{-4.2}
\newcommand{\ehjdaprm}{ 3.3}
\newcommand{\ecoswapr}{ 0.0024}

\newcommand{\fluxone}{45.1}
\newcommand{\efluxone}{2.3}
\newcommand{\magone}{9.49}

\newcommand{\fluxtwo}{28.5}
\newcommand{\efluxtwo}{1.4}
\newcommand{\magtwo}{9.50}

\newcommand{\fluxthr}{18.1}
\newcommand{\efluxthr}{0.9}
\newcommand{\magthr}{9.51}

\newcommand{\fluxfour}{11.7}
\newcommand{\efluxfour}{0.6}
\newcommand{\magfour}{9.35}

\slugcomment{Accepted for publication in The Astrophysical Journal}

\shorttitle{\objectname[NAME XO-1]{XO-1b} Thermal emission}
\shortauthors{Machalek et al.}

\begin{document}

%% LaTeX will automatically break titles if they run longer than
%% one line. However, you may use \\ to force a line break if
%% you desire.

\title{Thermal Emission of Exoplanet XO-1b}

%% Use \author, \affil, and the \and command to format
%% author and affiliation information.
%% Note that \email has replaced the old \authoremail command
%% from AASTeX v4.0. You can use \email to mark an email address
%% anywhere in the paper, not just in the front matter.
%% As in the title, use \\ to force line breaks.

\author{
Pavel~Machalek\altaffilmark{1,2},
Peter~R.~McCullough\altaffilmark{2},
Christopher~J.~Burke\altaffilmark{2},
Jeff~A.~Valenti\altaffilmark{2},
Adam~Burrows\altaffilmark{3},
%J.~E.~Stys\altaffilmark{2},
%R.~ Gilliland\altaffilmark{2},
%Christopher~M.~Johns-Krull\altaffilmark{4},
%Kenneth~A.~Janes\altaffilmark{5},
Joseph~L.~Hora\altaffilmark{4}
}

\email{pavel@jhu.edu}
\altaffiltext{1}{Department of Physics and Astronomy, Johns Hopkins University, 3400 North Charles St., Baltimore MD 21218}
\altaffiltext{2}{Space Telescope Science Institute, 3700 San Martin Dr., Baltimore MD 21218}
\altaffiltext{3}{Department of Astrophysical Sciences, Princeton University, Princeton, NJ 08544}
%\altaffiltext{4}{Dept. of Physics and Astronomy, Rice University, 6100 Main Street, MS-108, Houston, TX 77005}
%\altaffiltext{5}{Boston University, Astronomy Dept., 725 Commonwealth Ave.,Boston, MA 02215}
\altaffiltext{4}{Harvard-Smithsonian Center for Astrophysics, 60 Garden St., MS-65, Cambridge, MA 02138} 

%\altaffiltext{1}{Visiting Astronomer, Cerro Tololo Inter-American Observatory.
%CTIO is operated by AURA, Inc.\ under contract to the National Science
%Foundation.}
%\slugcomment{Submitted for publication in the Astrophysical Journal}

\begin{abstract}
We estimate flux ratios of the extrasolar planet XO-1b to its host star \objectname[NAME XO-1]{XO-1}~at 3.6, 4.5, 5.8 and 8.0 microns with the IRAC on the \spit~to be \flone~$\pm$~\eflone, \fltwo~$\pm$~\efltwo, \flthr~$\pm$~\eflthr~and ~\flfour~$\pm$~\eflfour, respectively.  The fluxes are inconsistent with a canonical cloudless model for the thermal emission from a planet and suggest an atmosphere with a thermal inversion layer and a possible stratospheric absorber. 
A newly emerging correlation between the presence of a thermal inversion layer in the planetary atmosphere and stellar insolation of the planet \citep{burr07b} is refined. The sub-stellar point flux from the parent star at XO-1b of $\sim$ 0.49 $\times$ 10$^9$ erg cm $^{-2}$ s $^{-1}$ sets a new lower limit for the occurrence of a thermal inversion in a planetary atmosphere.    
\end{abstract}

%% Keywords should appear after the \end{abstract} command. The uncommented
%% example has been keyed in ApJ style. See the instructions to authors
%% for the journal to which you are submitting your paper to determine
%% what keyword punctuation is appropriate.


%\objectname[NAME XO-1b]{XO-1b~}\\
%\objectname{2MASS J16021184+2810105}\\

%% Note that for sources with brackets in their names, e.g. [WEG2004] 14h-090,
%% the brackets must be escaped with backslashes when used in the first
%% square-bracket argument, for instance, \object[\[WEG2004\] 14h-090]{90}).
%%  Otherwise, LaTeX will issue an error. 
%\textit{revised by A. Burrows Feb 1 2008}\\
%\textit{revised by C. Burke Feb 12 2008}\\
%\textit{revised by P McCullough Feb 14 2008}\\
%\textit{v2 by A Burrows 3-11-2008}

\section{Introduction}
Over 270 extrasolar planets have been reported, more than 30 of which transit their primary star\footnote{\url{http://www.inscience.ch/transits/}}. In addition to the mass, radius and inclination of the planet evident from transits, atmospheric composition can also be studied through transmission spectroscopy, leading to detections of sodium, water and methane~\citep{charb02,tin07,swain08}.  
In addition, secondary eclipse observations provide broadband emission spectra \citep{knutson07b,charb08}, planetary brightness temperatures~\citep{charb05,dem05,har07} and even day-night  temperature contrast \citep{knutson07}. \cite{torres08} provide a reanalysis of light curves and RV measurements of all then known transiting planets. \\

For ``hot Jupiters'', planets with P$_{orb}$ $\lesssim$ 10 days, a favorable planet-star ratio in the IR allows for direct detection of the planetary atmosphere by comparing the combined flux from the star and the planet during and out of secondary eclipse at the superior conjunction. The contrast ratio in the mid-IR (1-10 microns) can be higher than 10$^{-3}$ (cf. observations of \objectname{HD 189733b} by \cite{charb08}) and theoretical predictions by \citet{burr06}, \citet{fort06} and \citet{burr07}, allowing for detection of fluxes from planets at secondary eclipse using the IRAC, IRS, and MIPS cameras of the \spit~(\citet{werner04}). Five planets have had their secondary eclipse fluxes measured in one or more IRAC bands: \objectname[NAME TrES-1b]{TrES-1} \citep{charb05}, \objectname{HD 209458b} \citep{dem05,knutson07b}, \objectname{HD 189733b} \citep{knutson07}, \objectname{HD 149026b} \citep{har07} and \objectname{GJ 436b} \citep{dem07}. In addition low resolution spectra of 2 transiting planets were obtained with the IRS spectrometer between $\sim$ 7 and 15 $\mu m$: \objectname{HD 189733b} \citep{grill07} and \objectname{HD 209458b} \citep{rich07}. \\

Recently a detection of an atmospheric feature attributed to water has been claimed by  \citet{tin07} and \citet{barman07} by studying the transit flux ratios of \objectname{HD 189733b} and \objectname{HD 209458b}, respectively. \citet{burr07} analyzed the secondary transmission spectra of \objectname{HD 209458b} at all 4 infrared IRAC Spitzer channels observed by \citet{knutson07b} and suggested the observations are consistent with an atmospheric thermal inversion layer and yet unknown stratospheric absorber. 
%The water absorption feature at around 4 $\mu m$ can be reversed into an \textit{emission} feature in the presence of an stratospheric absorber. 
A detailed study of the IR secondary eclipse planetary spectra of \objectname{HD 209458b}, \objectname{HD 189733b}, \objectname[NAME TrES-1b]{TrES-1}, \objectname{HD 149026b} and non-eclipsing \objectname{HD 179949b}, and \objectname[* ups And b]{$\upsilon$ And b} by \citet{burr07b} suggests that the presence of such a stratospheric absorber might be dependent on the flux from the star at the sub-stellar point on the planet as well as second order effects like metallicity and planetary surface gravity. In the \citet{burr07b} interpretation planets with high sub-stellar point flux (e.g., \objectname{HD 209458b}, OGLE-Tr-56b, OGLE-Tr-132b, \objectname[GSC 03549-02811]{TrES-2b} and \objectname[GSC 03727-01064]{XO-3b}) would have a stratospheric layer and a water feature in emission while planets with lower fluxes (\objectname[NAME XO-1b]{XO-1b},~\objectname[NAME TrES-1b]{TrES-1},~\objectname[TYC 3413-5-1]{XO-2b} and \objectname{HD 189733b}) would have no such layer and a water feature in absorption. \citet{fort07b} also suggest a similar division of planetary spectra based on incident stellar flux.
% and propose to call the close-in giant planets, which have high sub-stellar flux from the parent star and the associated atmospheric thermal inversions and anomalously bright mid-IR flux, the ``pM Class'' of planets. In contrast, planets with lower sub-stellar incident flux (``pL Class'') will absorb stellar flux deeper in their atmospheres and will not exhibit thermal inversions in their atmospheres. 
Based on the planetary sub-stellar point flux from the star, both \citet{burr07b} and \citet{fort07b} predict that \objectname[NAME XO-1b]{XO-1b}~ should not exhibit a thermal inversion in its atmosphere.    \\

We present observations of the infrared spectral energy distribution (SED) of the planet \objectname[NAME XO-1b]{XO-1b}~  \citep{xo1} in all 4 IRAC channels obtained during secondary eclipses with the IRAC of \spit. By comparing our  $\sim$ 4-8 $\mu m$ SED with atmospheric models, we test for the presence of a thermal inversion layer in \objectname[NAME XO-1b]{XO-1b}.

\section{Observations}

The InfraRed Array Camera \citep[IRAC;][]{fazio04} has a field of view of
5.2$\arcmin$ $\times$ 5.2$\arcmin$ in each of its four bands. Two adjacent
fields are imaged in pairs (3.6 and 5.8 microns; 4.5 and 8.0 microns). The
detector arrays each measure 256 $\times$ 256 pixels, with a pixel size of
approximately 1.22$\arcsec$ $\times$ 1.22$\arcsec$. We have observed
\objectname[NAME XO-1]{XO-1}~in all 4 channels in two separate Astronomical
Observing Requests (AORs) in two different sessions: the 4.5 and 8.0 micron
channels for 5.9 hours on UT 2007 April 02 (AOR 21374464) and the 3.6 and
5.8 micron channels for 5.9 hours on UT 2007 Aug 10 (AOR 21374208). We used
the 12 s frame time, obtaining 1620 full-array images in each bandpass with
a cadence of 13.2 s and an effective integration time of 10.4 s . The
pointing was not dithered and was selected such that for the Apr 2007
observations in the 4.5 and 8.0 micron channels, two bright calibrators for
\objectname[NAME XO-1]{XO-1}~(\objectname{2MASS J16021184+2810105}
J=9.939 J - K =0.412) were used: \objectname{2MASS J16021795+2813328}: J =
9.913 J - K = 0.564 and \objectname{2MASS J16020133+2809268}: J = 12.542 J
- K = 0.795.  The Aug 2007 observations in the 3.6 and 5.8 micron channels
used 2 bright calibrators: \objectname{2MASS J16021311+2809004}: J=11.045 J
- K = 0.652 and \objectname{2MASS J16020133+2809268}:  J = 12.542 J - K =
0.795.

We used the standard IRAC Basic Calibrated Data products (version 16.1)
described in the Spitzer Data
Handbook\footnote{\url{http://ssc.spitzer.caltech.edu/irac/dh/}},
which includes dark frame subtraction, multiplexer bleed correction,
detector linearization, and flat-fielding of the images.  We converted the
times recorded by the spacecraft in the FITS file header keyword DATE-OBS
to heliocentric Julian dates using the orbital ephemeris of the spacecraft
provided by the Horizons Ephemeris
System\footnote{\url{http://ssd.jpl.nasa.gov/}}.

%To account for fractional pixels in the relatively small circular aperture we have resampled the images in all four channels to a 10 times finer grid in each spatial direction using flux-conserving bilinear interpolation.

Prior to performing aperture photometry, we resampled the images in all four channels to a 10
times finer grid in each spatial direction using flux-conserving bilinear interpolation (similar
to \citet{har07}). With the implementation for aperture photometry that 
we used, resampling makes a marginal improvement in the photometry, i.e. a slightly lower r.m.s. of 
out-of-eclipse points, presumably related to how the routine handles fractional pixels at the edge of the 
aperture.
%Due to PSF under-sampling \citep{har07} in order to reduce pixelation
%noise during photometry we have resampled all images in all four channels
%to a 10 times finer grid in each spatial direction using flux-conserving
%bilinear interpolation. 
%The position of the two calibrators close to the
%image edge ($\leq$ 10 pixels) prevented us from performing sky subtraction
%by calculating the average of the entire sky annulus. 

The zodiacal
background was subtracted in each channel by constructing a histogram of
all pixels in each image and fitting a Gaussian to the distribution of the
zodiacal background brightness. A constant mean value of the Gaussian was
then subtracted from each pixel in the image to construct a
background-subtracted image.  The centroids of \objectname[NAMEXO-1]{XO-1}~and
the 2 calibrators were evaluated by fitting a Gaussian to the stellar flux
distribution. The pointing varied by 0.3 pixel, and the shifting of the
stellar centroid within a pixel, which have sub-pixel sensitivity
variations, resulted in a modulation of the stellar flux in the 3.6 and 4.5
micron channels (described below).\\

Aperture photometry was then performed on the images with an aperture
radius of 4 pixels, which was found to be the optimum value for all 4
channels. The size of the aperture was determined by minimizing the rms
scatter in the light curve for observations outside of the eclipse.
Apertures smaller than 4 pixels contained insufficient stellar flux and
larger apertures were more contaminated by the sky (especially in the high
background signal of the 5.8 and 8.0 micron channels). An appropriate
aperture correction for each channel was applied to the stellar flux value
according to the Spitzer Data Handbook of [1.112, 1.113, 1.125, 1.218] for
the [3.6; 4.5; 5.8 and 8.0] micron channels, respectively. The $\sigma$ of
the out-of-eclipse points was calculated iteratively using 3-$\sigma$
outlier rejection at each step until no more points were rejected. 
To remove cosmic rays the
resultant robust $\sigma$ was used to reject entire images which contain the 3-$\sigma$ outliers above and below the mean of the light curve. 1.4\%; 1.9\%; 3.0 \%, 2.4 \% of images from the 3.6, 4.5, 5.8 and 8.0 micron channels, respectively, were removed in this fashion. The higher rejection rate in the two redder channels is consistent with a higher number
of cosmic ray-affected pixels in these channels \citep{patten04}.
Throughout the analysis we have preserved flux units.

\subsection{3.6 and 4.5 micron time series}

\begin{center}
\begin{figure}[thp]
\centering
\includegraphics[scale=0.6]{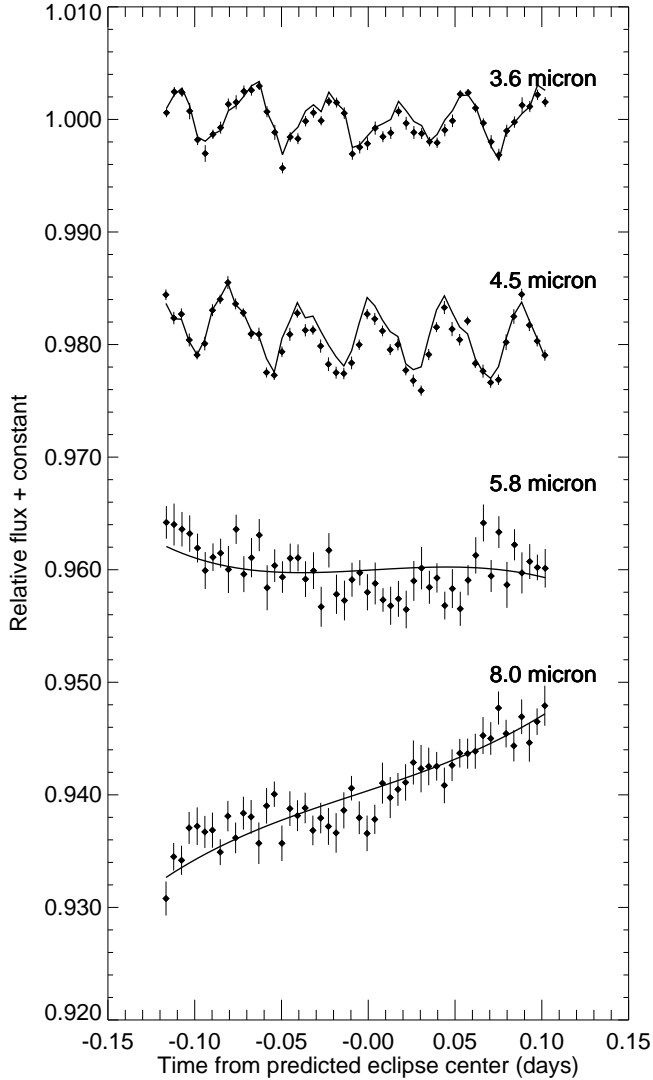}
\caption{\small Secondary eclipse observations of \protect\objectname{XO-1b}~ with 
IRAC on \spit~obtained on UT 2007 April 02 and UT 2007 Aug 10 in 3.6, 4.5, 
5.8 and 8.0 micron channels (from top to bottom) binned in 6-minute interval 
and normalized to 1 and offset by 0.02 for clarity. The overplotted solid 
lines do \textit{not} represent a fit to the data, but rather show the 
correction for the detector effects. The 3.6 and 4.6 micron time series 
are decorelated using \protect\objectname{XO-1} out-of-eclipse points and 
the 5.8 and 8.0 micron time series is detrended using a fit to a calibrator 
star in the field (see \S \ref{insb} \& \S \ref{sias} for details.)}
\label{fig:instru}
\end{figure}
\end{center}
\label{insb}

The 3.6-$\mu$m time series exhibited a sharp increase during the first $\sim$30 minutes of exposure for \objectname[NAME XO-1]{XO-1}~and the 2 calibrators, presumably  as a result of the instrument reaching a new equilibrium after previous observations. Such relaxation effects can reach several percent and usually stabilize within the first hour of observations of a new target. We have ignored the first 125 points ($\sim$30 min) in the 3.6-$\mu$m time series in addition to the high-sigma outlier rejection as described in the previous section.

A strong correlation between the sub-pixel centroid and stellar brightness was observed in both the 3.6-$\mu$m and 4.5-$\mu$m channels, with flux magnitudes of $\sim$ 0.6\% and 0.8\%, respectively. This well studied effect \citep{charb05,morales06} is due to the InSb detector intrapixel sensitivity variations as the spacecraft jitters $\sim$0.3 arcsec in orientation over a period of $\sim$3000 seconds \footnote{\url{http://ssc.spitzer.caltech.edu/documents/exoplanetmemo.txt}}. The uncorrected sub-pixel intensity variations are clearly visible in the time series of \objectname[NAME XO-1]{XO-1}~in the 3.6-$\mu$m and 4.5-$\mu$m channels in Fig.~\ref{fig:instru}. We have corrected for this sub-pixel intensity variations after \citet{charb08} by fitting a quadratic function to the photometric flux points of \objectname[NAME XO-1b]{XO-1b}~observed out-of-eclipse as a function of the x and y sub-pixel centroids: 

\begin{equation}
  \label{eq:subpixel}I_{subpixel}=~ b_{1} +~b_{2} \times~x~ +~b_{3} \times~x^{2} +~b_{4} \times~y~+~b_{5}~\times~y^{2}, 
\end{equation}
where $x$ and $y$ are the subpixel centroids of center of light of the star and $b_{n}$ are the fit parameters.
The rms residual of the \objectname[NAME XO-1]{XO-1}~time series for points outside of the eclipse after correction for the sub-pixel intensity variation was 0.0020, which is 18\% higher than a theoretical estimate of \objectname[NAME XO-1]{XO-1}~Poisson noise based on detector read noise and background noise. The time series was then normalized by taking the robust average of out-of-eclipse points and binned in 6-minute intervals containing approximately 30 individual measurements each (see Fig.~\ref{fig:fit}).

The 4.5-$\mu$m time series also exhibited an initial relaxation-induced brightness increase and consequently 139 points
corresponding to the first $\sim$30 minutes of observations were rejected, which is more points than in the 3.6-$\mu$m time series. The analysis of the time series was identical to that of the 3.6-$\mu$m time series.  The rms of out-of-eclipse points was 0.0024, which is 19\% higher than the theoretical estimate and is similar to that of \objectname[NAME TrES-1]{TrES-1} (rms=0.0027 \citet{charb05}).

We have tested for the linearity of the detector response in the 3.6 and 4.5~micron channels in which XO-1 is close to the onset of detector non-linear response. Using a subset of data from the SAGE survey \citep{meix06} obtained in the high dynamic range (HDR) mode of IRAC camera with both 0.6s and 12.0s integration times we are able to determine that both the 3.6~micron and 4.5~micron XO-1 fluxes are unsaturated and in the detector linear regime response. Table \ref{flux} shows the absolute XO-1 fluxes and the instrumental magnitudes in the four IRAC channels. 
%We encourage other observers of secondary eclipses with IRAC to quote their absolute stellar fluxes and test for detector non-linearity response. 

\begin{deluxetable}{ccc}
\tablecolumns{3}
\tablewidth{0pt}
\tablecaption{XO-1 absolute fluxes }
\tablehead{
\colhead{IRAC channel}  &\colhead {XO-1 flux}&\colhead{XO-1 instrumental}\\
\colhead{effective $\lambda$}  &\colhead {}&\colhead{magnitude}\\
\colhead{(microns)}  &\colhead {(mJy)} &\colhead {(mag)}}
\startdata
3.6  & \fluxone~$\pm$~\efluxone &\magone\\
4.5  & \fluxtwo~$\pm$~\efluxtwo &\magtwo\\
5.8  & \fluxthr~$\pm$~\efluxthr &\magthr\\
8.0  & \fluxfour~$\pm$~\efluxfour &\magfour\\
\enddata
\label{flux}
\end{deluxetable}

\subsection{5.8 and 8.0 micron time series}
\label{sias}
\begin{figure}[tph]
\centering
\includegraphics[totalheight=0.8\textwidth]{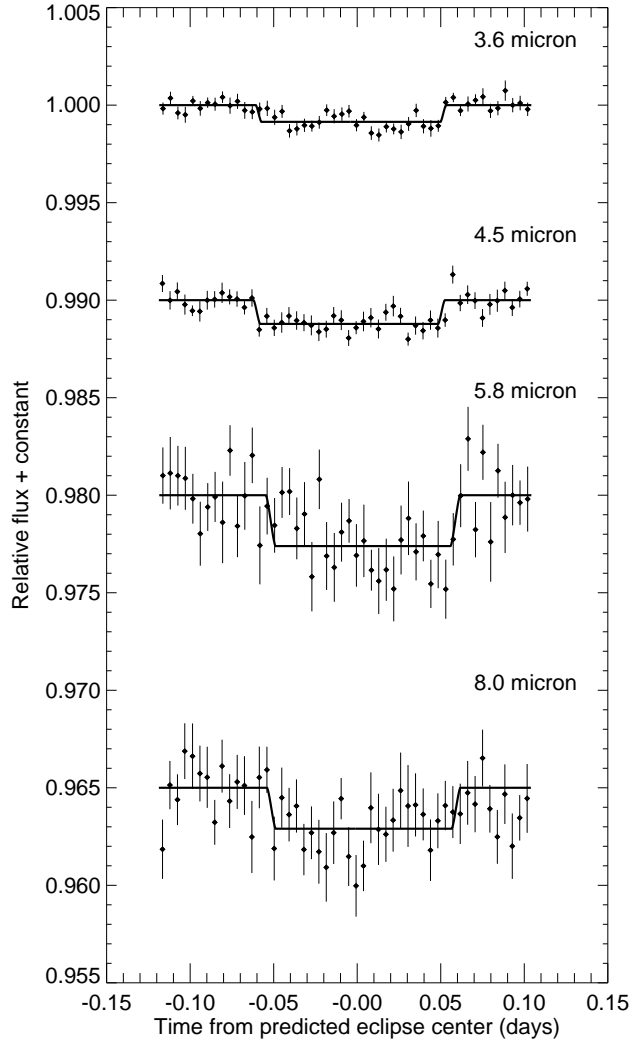}
\caption{Secondary eclipse of \protect\objectname{XO-1b} observed with IRAC on \spit~in 3.6, 4.5, 5.8, and 8.0 micron channels (top to bottom) corrected for detector effects, normalized and binned in 6-minute intervals and offset for clarity. The best-fit eclipse curves are overplotted. }
\label{fig:fit}
\end{figure}

The 5.8 and 8.0 micron time series were recorded with Si:As detectors and do not thus exhibit the prominent sub-pixel intensity variations evident in the 3.6~micron and 4.5~micron channels. The first $\sim$30 minutes of observations (139 data points) were rejected as the instrument settled into a new equilibrium state.  Fig. ~\ref{fig:instru} shows intensity variation with time, which is caused by changes in the effective gain of individual pixels over time. 
%This effect has been observed before by \citet{dem05}, both in the IRAC camera and in the IRS and MIPS 24~micron cameras and is dependent on the illumination level of the individual pixel \citep{knutson07,knutson07b}. 
This effect has also been been observed by \citet{har07} at 8~micron with IRAC and by \citet{dem06} at 16~micron with IRS. The intensity variations are dependent on the illumination level of the individual pixel \citep{knutson07,knutson07b}, pixels with high illumination will reach their equilibrium within $\sim$1 hour, but lower illumination pixels increase in intensity over time, approximately proportional to the inverse of the logarithm of illumination. We have decided not to correct each pixel in the image as \citet{knutson07} have done in their 33-hour observation of HD 189733 in the 8~micron  channel of IRAC and instead fit a combined linear and quadratic logarithm function of time from the beginning of observations to the time series of the bright calibrator \objectname{2MASS 16020133+2809268}. The fit to the time series of the calibrator was then used to remove the detector ramp from the time series of \objectname[NAME XO-1]{XO-1}~as depicted in Fig.~\ref{fig:instru}:

\begin{equation}
  \label{eq:ramp}I_{model} = a_{1} + a_{2}  \times~ \Delta t + a_{3} \times~  ln\Delta t + a_{4} \times~ (ln\Delta t)^{2},
\end{equation}
where $I_{model}$ is the model flux, $\Delta t$ is the time since the beginning of observations and $a_{i}$ are the free parameters. The detector ramp intensity decreased in flux during the 5 hours of observation by $\sim$ 0.2 \%, following a similar trend seen by \citet{knutson07b} in their 5.8~micron time series of brighter \objectname{HD 209458b}. After removal of the detector ramp and normalization, the rms of unbinned 5.8 micron out-of-eclipse points for \objectname[NAME XO-1]{XO-1}~was 0.0081, which is 44\% higher than theoretical Poisson noise, based on detector read noise and background noise.\footnote{\url{http://ssc.spitzer.caltech.edu/documents/irac\_memo.txt}} 
%The internal scattering of photons inside the Si:As arrays is likely  responsible for the fact that we do not approach the Poisson limit as closely as in the 3.6~micron and 4.6~micron channels. 
The normalized and binned  5.8~micron time series is depicted in Fig.~\ref{fig:fit}.

Fitting the detector ramp using a calibrator star, which is 
non-variable at the 0.012 level after detector ramp removal, allows us 
to bypass using XO-1 itself to remove the detector ramp by making a fit 
to its 5.8~micron out-of-eclipse points. The choice whether to correct 
for the detector ramp of XO-1 using the photometry of a calibrator star 
or the target star itself could be a limiting factor in our analysis.  
Deming (2008, p.c.) has indicated that in a long ($>$ 10 hours) series of 
full-frame 12s-exposure IRAC photometry, fitting the detector ramp to 
the target star while masking out the points in the eclipse could be 
more appropriate.  We implemented each of the two alternate methods of 
correction, and for these observations of XO-1b, the measured depths at 
5.8~micron are within 1-$\sigma$ of each other.

%Fitting the detector ramp using a calibrator star, which is non-variable at the 0.012 level after detector ramp removal, allows us to bypass the problems of using \objectname[NAME XO-1]{XO-1}~itself to remove the detector ramp by making a fit to its  5.8~micron out-of-eclipse points\footnote{As a test, we have used the out-of-eclipse points of the \objectname[NAME XO-1]{XO-1}~target star for fitting the detector ramp. Our secondary eclipse depths in the decorrelated 5.8 \& 8.0 micron time series changed by less than  1-$\sigma$.}. The choice whether to correct for the detector ramp of XO-1 using the photometry of a calibrator star or the target star itself is a limiting factor in our analysis and there are indicators that in long duration ($>$ 10 hours) full frame 12s exposure time IRAC  photometry fitting the detector ramp to the target star while masking out the points in the eclipse might be more appropriate (Deming 2008, private communication). 
%The out-of-eclipse detector ramp technique, where the detector ramp fitting is applied to the target star time series, is adopted when one observes bright stars in sub-array mode (\cite{knutson07,har07}), which has no calibrators in the field. 

\citet{charb05}, \citet{knutson07}, \citet{har07} and \citet{dem07} have reported a nonlinear flux increase over time in the 8.0~micron
IRAC channel. We also detect a nonlinear increase in the brightness of \objectname[NAME XO-1]{XO-1}~and in the 2 calibrators, in addition to the sharp increase during the first $\sim$30 minutes of observations (137 points). The initial ramp-up data were discarded and the $\sim$ 1.6\% non-linear increase of the \objectname[NAME XO-1]{XO-1}~time series over 5 hours (see Fig.~\ref{fig:instru}) was removed by fitting a combined linear and quadratic logarithm function (eq. \ref{eq:ramp}) to the time series of the first calibrator \objectname{2MASS J16021795+2813328} and dividing the \objectname[NAME XO-1]{XO-1}~time series by the fit. The time series was normalized using out-of-eclipse points and binned into 6-minute bins (Fig.~\ref{fig:fit}) for viewing clarity.
The resulting rms of unbinned out-of-eclipse points for the \objectname[NAME XO-1]{XO-1}~time series was 0.0075, 74\% above Poisson noise, but similar to that of \objectname[NAME TrES-1b]{TrES-1} (rms = 0.0085, \citep{charb05}).

\section{Analysis}
\label{anl}
We fit the secondary eclipse light curves using the formalism of \citet{agol02} with no stellar limb darkening and adopt stellar and orbital parameters \citep{holman06}\footnote{We have also reduced the data using the stellar and orbital parameters from \citet{torres08} as a test, but the eclipse depths changed negligibly and eclipse mid-center timings were all within 1-$\sigma$.}: R$_{\star}$ = \rs $^{\ersp}_{\ersm}$ R$_{\sun}$, R$_{p}$ = \rp $^{\erpp}_{\erpm}$ R$_{Jup}$\footnote{1 R$_{Jup}$ = 71,492 km.}, $i$ =~89.31$^{+0.46}_{-0.53}$ degrees, and $a$ = \semi~$\pm$~\esemi~AU with ephemeris \citep{xo1}

\begin{equation}
  \label{eq:ephm} T_{c}(E) = 2,453,808.9170(HJD) + E(3.941534~days) \, .
\end{equation}
We fit the depth of the eclipse $\Delta F$ and the timing of the centroid $\Delta T$ independently in all 4 channels in the unbinned light series using Levenberg-Marquardt minimization \citep{press92} with an equal error assigned to all points, which is equal to the rms of out-of-eclipse points in each time series. Best-fit eclipse curves are plotted in Fig.~\ref{fig:fit} and the eclipse parameters are listed in Table \ref{tbl1}. They are the channel wavelength, eclipse depth $\Delta F$, eclipse mid-center time in HJD, and the timing offset  $\Delta t$ in minutes from the expected secondary eclipse mid-center time for an assumed eccentricity of zero, and the reduced $\chi ^{2}$. The reduced $\chi ^{2}$ is close to 1.0 in all 4 channels, indicating a good fit to the data.

\begin{deluxetable*}{ccccc}
\tablecolumns{5}
\tablewidth{0.8\textwidth}
\tablecaption{Secondary eclipse best fit parameters }
\tablehead{
\colhead{$\lambda$}  &\colhead{Eclipse Depth $\Delta F$} &\colhead{Eclipse Center Time} &\colhead{Time offset $\Delta T$ } & \colhead{Reduced $\chi^{2}$}\\
\colhead{(microns)}  &\colhead{} &\colhead{(HJD)} &\colhead{(min)} & \colhead{}}
\startdata
3.6  &\flone~$\pm$ \eflone & \hjdone~$\pm$ \ehjdone& \dtone~$\pm$ \edtone&  \chione\\
4.5  & \fltwo~$\pm$ \efltwo & \hjdtwo~$\pm$ \ehjdtwo& \dttwo~$\pm$ \edttwo&  \chitwo\\
5.8  & \flthr~$\pm$ \eflthr & \hjdthr~$\pm$ \ehjdthr& \dtthr~$\pm$ \edtthr&  \chithr\\
8.0  & \flfour~$\pm$ \eflfour & \hjdfour~$\pm$ \ehjdfour& \dtfour~$\pm$ \edtfour&  \chifour\\
\enddata
\label{tbl1}
\end{deluxetable*}

To estimate the errors on the depth and mid-eclipse timing we performed the error analysis using the bootstrap method from \citet{press92}. The bootstrap method makes no prior assumptions about the distribution of the noise in the data and the data points are not altered as in the Monte-Carlo analysis. For 10,000 trial runs we have randomly drawn  with replacement points from the normalized- and detector-effect-decorrelated, but otherwise unaltered, light curve, until we had the same number of data points in the light curve that we started with. During each iteration we performed the full eclipse fitting for eclipse depth $\Delta F$ and eclipse mid-center $\Delta T$.  The 1$-\sigma$ errors for $\Delta F$ and $\Delta T$ were computed by fitting a Gaussian to the respective 1-D distribution of bootstrap points and are reported in Table \ref{tbl1}.  \\

The eclipse depth errors $\Delta F$ = [\eflone, \efltwo, \eflthr, \eflfour] for the J = 9.939 \objectname[NAME XO-1]{XO-1}~compare favorably with the eclipse depth errors [0.00009, 0.00015, 0.00043, 0.00026] of \citet{knutson07b} for the J = 6.591 HD 209458, despite the fact that \objectname[NAME XO-1]{XO-1}~is dimmer. \objectname[NAME XO-1]{XO-1}~observations were made in the full-array mode with 10.4-s integration time and readout time 2.8 s for a total of 1,620 images, while the \objectname{HD 209458} observations were made in sub-array mode with exposure time of 0.1 s in sets of 64 in each channel totaling 26,880 usable images. The S/N scales as~$\propto$ $\sqrt{n_{exp}}$ * $f_{e}$  * $\Delta t$ / $\sigma_{total}$, where $n_{exp}$ is the number of exposures during the duration of the eclipse,~$f_{e}$ is the stellar flux in signal electrons,~$\Delta t$ is the integration time and $\sigma_{total}$ is the combined Poisson, readout and background noise. The predicted eclipse depth errors  for \objectname[NAME XO-1]{XO-1}  in the four IRAC channels are then [0.4; 0.5; 0.9; 2.5] times the respective eclipse depth errors for the  21 times brighter \objectname{HD 209458}.  \\

To test the robustness of our data reduction and analysis technique
and consistency with other observations in the IRAC full-array mode we
have re-reduced the 4.5 and 8.0~micron IRAC secondary eclipse
data of \objectname[NAME TrES-1b]{TrES-1} by \citet{charb05} with our pipeline, rejecting the
first 30 minutes in both channels. The procedure described in \S\ref{insb} and \S\ref{sias} was used together with updated stellar and
planetary parameters of \citet{torres08} to derive the eclipse depths
relative to the star \objectname[NAME TrES-1b]{TrES-1} of $\Delta F_{4.5\mu m}$ = 0.00043 and $\Delta F_{8.0\mu m}$ = 0.00194, which are -1.1$\sigma$ and -0.9$\sigma$
away from the \citet{charb05} secondary eclipse depths of $\Delta F_{4.5\mu m}$ = 0.00057 $\pm$ 0.00013 and $\Delta F_{8.0\mu m}$ = 0.00225 $\pm$ 0.00036, respectively.  Our secondary eclipse [4.5; 8.0]micron channel mid-center timing offsets of [+12.5; -2.3] minutes are [-1.1$\sigma$; -2.0$\sigma$] away from the
\citet{charb05} [4.5; 8.0]micron channel values +19.6 $\pm$ 6.6 min and +8.3 $\pm$ 5.2 minutes, respectively. While the eclipse depths are consistent at the $\sim$ 1-$\sigma$ level, the eclipse mid-center timing offset in the 8.0~micron channel is only mildly consistent at the 2.0$\sigma$ level, probably because \citet{charb05} observations were made using multiple Astronomical Observation Requests (AORs) which resulted in arbitrary flux shifts in the time series. \citet{charb05} do not mention how they have corrected for these flux shifts. Despite this fact our pipeline is capable of reproducing their results to within $\sim$ 1-$\sigma$ in eclipse depth and $\sim$ 2-$\sigma$ in mid-eclipse timing.      
%as \citet{charb05} used a third-order polynomial fitted to a calibrator star to remove the baseline trend in brightness from the time series of TrES-1 in the 8.0micron channel while we use the functional form in Eq. \ref{eq:ramp},which could result in shift in the eclipse mid-center timings as ingress and egress in the IR last a short duration and are poorly constrained in our data. 
We have thus demonstrated that our reduction pipeline is robust and the secondary eclipse depth estimates are  consistent with other major full-array pipelines.  \\    

%The 1-$\sigma$ error on the depths $\Delat F$ in each channel were estimated from the distribution of the MC best fit depths and are reported in Table \ref{tbl1}\\
%The predicted S/N ratios based on the predicted flux densities of XO-1 in the 4 channels are depicted in Tab. \ref{tb:sn}, together with the predicted and observed rms of the final light curve.          

\section{Discussion}
 The eclipse mid-center timings for \objectname[NAME XO-1b]{XO-1b}~ in Table \ref{tbl1} are individually  consistent with zero timing residuals for a circular orbit based on the ephemeris by \citet{xo1}, but the April and August 2007 combined timings show a time shift. The UT Apr 2 2007 observations of 4.5~micron and 8.0~micron channels have a combined eclipse mid-center timing of \hjdapr ~$\pm$ \ehjdapr~HJD, with a delay of \hjdaprm~$\pm$\ehjdaprm~min, while the UT 2007 Aug 10 observations of 3.6~micron and 5.8~micron channels have a  combined center of eclipse time of \hjdaug~$\pm$  \ehjdaug~HJD, with a delay of \hjdaugm~$\pm$\ehjdaugm~min. The mid-eclipse timing offsets can be interpreted as evidence for further planets in the system, as can a non-uniform brightness distribution of the planet (which can change the shape of eclipse ingress and egress and, thus, shift the eclipse mid-center time \citep{will06}), or as evidence for a non-zero eccentricity. Using the larger mid-eclipse April time offset, the approximate equation (e.g. \citet{kopal59} Eq. 9-23)
  
\begin{equation}
  \label{eq:ecc} e \times cos(\omega) \simeq \frac{\pi \Delta t}{2P}, 
\end{equation}
where $e$ is the eccentricity, $\omega$ is the longitude of periastron, $P$ is the orbital period, and $\Delta t$ is the centroid time shift from expected time of secondary eclipse, allows us to set a 2 $\sigma$ upper limit on e$\times$cos($\omega$)~$<$~\ecoswapr.

%\textbf{Calculate the combined time shift and asses whether it is evidence for non-zero eccentricity.}\\

\begin{figure}[tph]
\centering
\includegraphics[angle=90,totalheight=0.32\textwidth]{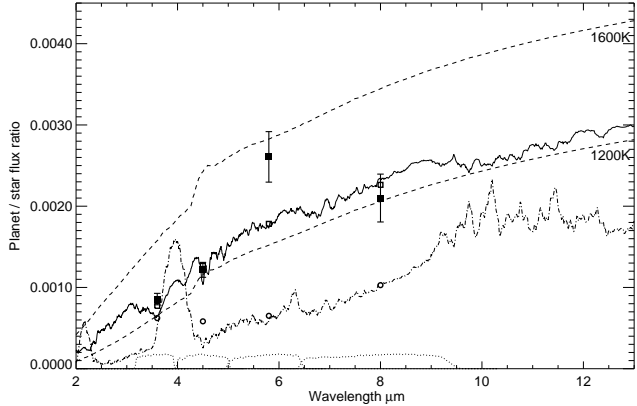}
\caption{~\spit~IRAC secondary eclipse depths for \protect\objectname{XO-1b} with bootstrap error bars (filled squares). The predicted emission spectrum of the planet with an upper atmospheric absorber of $\kappa _{e}$ = 0.1 cm$^{2}$/g and a redistribution parameter of P$_{n}$=0.3 is plotted as a solid line. A model with no atmospheric absorber and a redistribution parameter of P$_{n}$=0.3 is over plotted with dot-dashed line (see \S \ref{anl} for details). The band-averaged flux ratios are plotted as open squares and open circles for the models with and without stratospheric absorber, respectively. The normalized~\spit~IRAC response curves  for the 3.6-, 4.5-, 5.8-, and 8.0~micron channels are plotted at the bottom of the figure (doted lines). The theoretical flux ratios obtained from a \protect\objectname{XO-1} stellar spectrum (from \protect\url{http://kurucz.harvard.edu/stars/XO1/}) and an assumed black-body spectrum for the planet at [1200, 1600] K are plotted as dashed lines. }
\label{fig:atmo}
\end{figure}

The \objectname[NAME XO-1b]{XO-1b}~ eclipse depths in Fig.~\ref{fig:atmo} show several trends.
The planet-to-star contrast of \objectname[NAME XO-1b]{XO-1b}~ peaks in the 5.8 $\mu m$ channel, with a decrease
in the 3.6~micron and 4.5~micron channel and a slight decrease towards
the 8.0~micron channel. Furthermore the flux in the 4.5 $\mu m$
channel is higher than in the 3.6~micron channel, which does not match
the general character of the cloudless models of \citet{burr06} for redistribution parameter
P$_{n}$ =0.3 (dot-dashed line and open circles for band-averaged ratios in Fig.~\ref{fig:atmo}), which predict a lower flux at 4.5 microns than at 3.6 microns; the model with thermal inversion predicts the opposite. Since the observations manifest a higher flux in the 4.5 micron channel than at 3.5 micron, a thermal inversion in the atmosphere might be indicated.
%, which predict a water \textit{absorption} feature near 3.6~micron and thus a higher flux ratio in the 3.6~micron channel than in the 4.5~micron channel.  
P$_{n}$=0 corresponds to no heat redistribution from the planetary day-side to the night-side and P$_{n}$ = 0.5 stands for full redistribution (see \citet{burr07b} for details). 
The possibility of thermal inversion in a planetary atmosphere has been suggested by \citet{hubeny03},~\citet{burr06,burr07} and \citet{fort07b}. Recently, \citet{burr07,burr07b} suggested that model spectra could match the observations of \objectname{HD 209458b} and \objectname{HD 149026b} if a stratospheric absorber of unknown composition (possibly tholins, polyacetylenes, TiO or VO) were present in the atmosphere of the planet. The presence of a stratospheric absorber would yield a thermal inversion in the planetary atmosphere and the presence of the water features in emission for a variety of heat redistribution parameters P$_{n}$. \\

Our observations suggest the presence of a thermal inversion layer and a possible stratospheric absorber in the atmosphere of the \objectname[NAME XO-1b]{XO-1b}~planet. 
The solid line and open squares in Fig.~\ref{fig:atmo} depict an atmospheric model of \objectname[NAME XO-1b]{XO-1b}, following the methodology of \citet{burr06,burr07} with a thermal inversion and a stratospheric absorber of opacity of $\kappa _{e}$ = 0.1 cm$^{2}$/g and redistribution parameter of P$_{n}$ = 0.3. The latter model fits the data better than the canonical cloudless model with P$_{n}$ = 0.3 (dot-dashed curve and open circles for averaged band ratios).  The band-averaged flux ratios for the model with a stratospheric absorber (open squares) are within the error bars for the 3.6, 4.5, and 8.0~micron channels, but are inconsistent by 2.7$\sigma$ with the band-averaged flux ratios for the 5.8~micron channel. This is similar to the situation for the IRAC fit to the observations by HD 209458b by \citet{knutson07b}.  The absorber-free canonical model (dot-dashed line) is clearly inconsistent with our observations (Fig.~\ref{fig:atmo}) of XO-1b in all 4 channels by [3.4$\sigma$, 7.1$\sigma$, 6.3$\sigma$, 3.7$\sigma$], respectively.  \\

\begin{figure}[tph]
\centering
\includegraphics[angle=90,totalheight=0.32\textwidth]{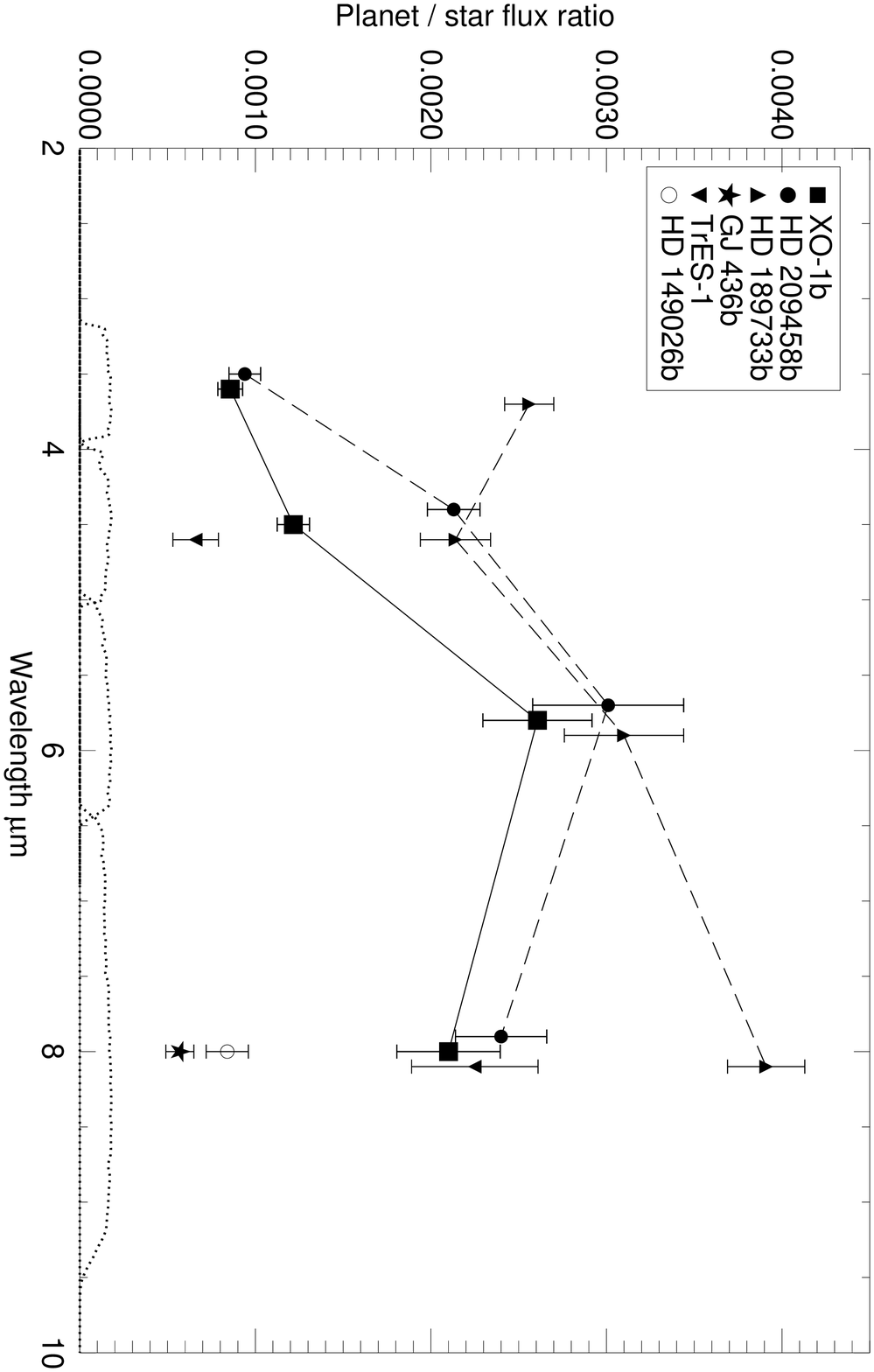}
\caption{Comparison of \spit~IRAC secondary eclipse depths: \protect\objectname{XO-1b} (filled square) from this paper; \protect\objectname{HD 209458b} (filled circle) from \citet{knutson07b}; \protect\objectname{HD 189733b} (filled triangle) from \citet{charb08}; \protect\objectname{GJ 436b} (filled star) from \citet{dem07}; \protect\objectname{TrES-1} (filled upside down triangle) from \citet{charb05}; and \protect\objectname{HD 149026b} (open circle) from \citet{har07}. The central wavelengths have been offset by [+0.1;-0.1]~microns for clarity. The normalized~\spit~IRAC response curves  for the 3.6-, 4.5-, 5.8-, and 8.0~micron channels are plotted at the bottom of the figure (doted lines). }
\label{fig:comp}
\end{figure}

 \citet{burr07b} and \citet{fort07b} suggested that the presence of the stratospheric absorber might be correlated with the incident flux from the star at the sub-stellar point on the planet, the precise level of which is yet to be refined. The presence of an irradiation-induced stratospheric absorber has been suggested by \citet{burr07} for \objectname{HD 209458b}  (see our Fig.~\ref{fig:comp}) with a sub-stellar flux of $\sim$1.07 $\times$ 10$^9$ erg cm $^{-2}$ s $^{-1}$ at a distance $a$ = 0.045 AU. Interestingly, \objectname[NAME XO-1b]{XO-1b}~has a lower sub-stellar flux of $\sim$ 0.49 $\times$ 10$^9$ erg cm $^{-2}$ s $^{-1}$ and a semi-major axis of $a$ = 0.0488 AU, but still manifests evidence for a thermal inversion. A recent study of the broadband infrared spectrum of \objectname{HD 189733b} (see our Fig.~\ref{fig:comp}) by \citet{charb08} finds no evidence for an atmospheric thermal inversion, despite a similar sub-stellar point flux of $\sim$ 0.47 $\times$ 10$^9$ erg cm $^{-2}$ s $^{-1}$  \citep{burr07b} with a smaller semi-major axis $a$ = 0.0313 AU.   Further study of planetary atmospheres should refine the concept of this sub-stellar flux boundary with respect to the presence/absence of a stratospheric absorber and thermal inversion. \\

Atmospheric water detection has been claimed in the transit broadband spectra of \objectname{HD 189733b} by \citet{tin07} and in its secondary eclipse spectra by \citet{fort07}.  \citet{burr07} also found evidence for  water vapor emission in the atmosphere of \objectname{HD 209458b}. Our data can be interpreted as evidence for rovibrational band of water emission longward of $\sim$4.0 microns, which manifests itself as a flux enhancement in Fig.~\ref{fig:atmo} compared to the cloudless model. The depth of the flux ratio trough near the 3.6~micron channel in the atmospheric models with stratospheric absorber strongly depends on the redistribution parameter P$_{n}$ (\citet{burr07b}, especially their Fig.~4). Further modeling would allow tighter constraints on the P$_{n}$, not just for \objectname[NAME XO-1b]{XO-1b}, but for a variety of planets. The fortuitous importance of the 3.6~micron IRAC channel to the study of planetary atmospheres is likely to be enhanced as the \spit~runs out of cryo-coolant in 2009, when only the 3.6~micron and 4.5~micron channels will be available.

\section{Conclusion}
We report the estimated flux ratios of the planet \objectname[NAME XO-1b]{XO-1b}~ in the \spit~IRAC 3.6, 4.5, 5.8 and 8.0-$\mu m$ channels. We find that the estimated fluxes are not consistent with a canonical cloudless model for thermal emission from the planet and instead may indicate an atmosphere with as yet unknown stratospheric absorber and a likely thermal inversion, which would cause the water band longward of 4.0 microns to switch from absorption to emission. The atmospheric model with a thermal inversion produces a tight match to the data at 3.6, 4.5, and 8.0 microns, but is inconsistent by 2.7$\sigma$ with observations at 5.8 microns.  This is similar to observations of HD 209458b \citep{knutson07b}.    

The presence or absence of the stratospheric absorber and thermal inversion layer has been linked to the flux from the parent star at the sub-stellar point on the planet. The \objectname[NAME XO-1b]{XO-1b}~ sub-stellar point flux of $\sim$0.49 $\times$ 10$^9$ erg cm $^{-2}$ s $^{-1}$ is the lowest so far reported for a planetary atmosphere with a thermal inversion. Observations of atmospheres of other planets may permit a better understanding of the thermal inversion layer and parametrization of the characteristics that create such a thermal inversion.   

\acknowledgments
The authors would like to thank J.~E.~Stys, R.~ Gilliland, C.~M.~Johns-Krull, and K.~A.~Janes for helpful discussions. The authors would also like to acknowledge the use of publicly available routines by Eric Agol and Levenberg-Marquardt least-squares minimization routine MPFITFUN  by Craig Markwardt.  P.M. and P.R.M. were supported by the Spitzer Science Center Grant C4030 to the Space Telescope Science Institute. A.B. was supported in part by NASA under grants NAG5-10760 and NNG04GL22G. This work is based on observations made with the Spitzer Space Telescope, which is operated by the Jet Propulsion Laboratory, California Institute of Technology under a contract with NASA. This publication also makes use of data products from the Two Micron All Sky Survey, which is a joint project of the University of Massachusetts and the Infrared Processing and Analysis Center/California Institute of Technology, funded by the National Aeronautics and Space Administration and the National Science Foundation. The authors would like to thank the reviewer Dr. Drake Deming for his helpful comments which have substantially improved the manuscript.  \\

%{\it Facilities:} \facility{Nickel}, \facility{HST (STIS)}, \facility{CXO (ASIS)}.
%{\it Facilities:} \facility{Spitzer (IRAC)}

%\clearpage

%\clearpage

%\clearpage

%\begin{figure}
%\epsscale{.80}
%%\plotone{f1.eps}
%\caption{Derived spectra for 3C138 \citep[see][]{heiles03}. Plots for all sources are available
%in the electronic edition of {\it The Astrophysical Journal}.\label{fig1}}
%\end{figure}

%\clearpage
%\centering

\end{document}